\documentclass[%
reprint,
 amsmath,amssymb,
 aps,
]{revtex4-2}

\usepackage{xcolor}
\usepackage{graphicx}
\usepackage{float}
\usepackage{siunitx}
\usepackage{makecell}
\usepackage{soul}
\usepackage[normalem]{ulem}   

\usepackage{hyperref}
\hypersetup{
    colorlinks=true,
    linkcolor=blue,
    citecolor=blue,
    filecolor=blue,
    urlcolor=blue
}

\newcommand{\La}{\mathcal{L}}

\begin{document}

\title{Cold Beam of $^7$Li$^4$He Dimers}
\author{Jeremy Glick, William Huntington, Daniel Heinzen }
\affiliation{Department of Physics, 
 University of Texas at Austin, Austin, TX 78712
}%

\date{\today}

\begin{abstract}
We demonstrate an approach for producing a high flux beam of $^7$Li$^4$He dimers by performing post-nozzle seeding of a lithium beam into a supersonic helium expansion. The molecular beam has a longitudinal temperature of 13(6) mK and a continuous flux on the order of $10^{11}$ dimers/s. Monte Carlo simulations of molecular formation based on ab-initio quantum scattering calculations are carried out and compared to the experimentally observed molecular flux. Extensions of this work could lead towards a more quantitative understanding of few-body collision processes of alkali and helium atoms. 

\end{abstract}
\maketitle

Methods to produce cold molecules are of interest for studies of molecular collisions and ultra-cold chemistry, precision measurement experiments, and quantum science \cite{Jun_2009_moleculeOverview,
Quéméner_2012_ultracoldMolecules, Jun_2017_molOverview}. In recent years, there has been great interest in molecules where fully quantum calculations are possible for studies of fundamental chemical processes such as three-body recombination \cite{Greene_2011_K3, Perez_2021_K3Theory, Kokkelmans_2021_K39, Rios_2022_three_body_overview, WangBin_2022_He_alkaline_theory}. Here, a great deal of experimental work has been carried out \cite{Thomas_1984_hyrogen, Silvera_1992_hydrogen, Toennies_2002_HeDimers,
Mazzoni_2000_RbMOT, Weber_2003_CsMOT, Doyle_2011_bufferGasFormationRates, Hickson_2011_beam, Weinstein_2017_TiHe, Weinstein_2012_reactionRate, Ye_2014_review,
Hecker_2022_K3, Hecker_2023_energyScaling}, and quantitative theoretical understanding of these measurements is an important goal. 

Van der Waals (vdW) molecules containing hydrogen or helium are well suited for theoretical studies given their relatively simple structure \cite{Kleinekath_1999_singleBoundState, Esry_2009_recombination, Esry_2011_hydrogen, Tiesinga_2020_HeLi_theory, Chang_2023_HeK3_Theory}. For example, all alkali-helium pairs are predicted to contain a single bound-ground state \cite{Kleinekath_1999_singleBoundState} allowing for calculation of fully quantum ab-initio rate constants. More complex molecules containing a multitude of recombination channels makes such calculations challenging although progress has been made in some systems \cite{ Greene_2012_Emifov, Hecker_2022_K3}. Creating these alkali-helium pairs is not simple as the binding energies are predicted to be on the order of single to tens of $k_{\rm b} \times$ mK \cite{Kleinekath_1999_singleBoundState, Tiesinga_2020_HeLi_theory}. New methods to produce and study these molecules are in important step for connecting ab-initio calculations to experiment.

An intense cold atomic beam source that we have recently developed \cite{Heinzen_2023_Cold_Atom_Source} can be used to produce alkali-helium dimers for studies of few-body processes. Using this source, we have produced a high flux beam of $^7$Li$^4$He dimers with a longitudinal temperature of 13(6) mK. We have also carried out Monte Carlo simulations of the molecular formation that incorporate currently available theoretical rate constants. Adopting our approach to other species opens up the possibility of further studies of van der Waals molecular formation and collisions in the 1 mK to 100 mK temperature range, intermediate between the sub-mK or above 100 mK temperatures of most previous work with cold molecules. 

Various techniques have been used to produce cold molecules. 
Magneto-optical traps and laser cooling have been applied to a number of species and can
achieve $\mu$K temperatures \cite{DeMille_2010_LaserCooling, Jun_2009_moleculeOverview}, but are not easily applicable to ground state helium . Buffer-gas based sources using helium have been widely used to create and cool a variety of molecules  \cite{Hutzler_2012_buffer_gas_beam, Doyle_2008_AgHe, Weinstein_2017_TiHe}, including the first observation of LiHe by Tariq \emph{et al.} \cite{Weinstein_2013_LiHe}. Limited by the vapor pressure of helium, temperatures are typically in the single Kelvin range \cite{Hutzler_2012_buffer_gas_beam}. The CRESU method, based on seeded supersonic helium expansions, is also typically in the single to tens of Kelvin range \cite{Hickson_2011_beam}.  

In our approach, we make use of highly expanded supersonic helium, which can reach mK temperatures \cite{Wang_1988_1_mK_jet}, enabling for example the first observation of He$_2$ dimers and He$_3$ trimers \cite{Toennies_1996_HeDimer}. While the standard approach for producing seeded jets has been to perform seeding pre-expansion or at the opening of the jet nozzle \cite{Smalley_1977_molecular_spectroscopy, Dietz_1981_Smalley_Source, Hillenkamp_2003_condensation_limited_cooling, Duncan_2012_Smalley_source_review, Aggarwal_2021_Seeded_SrF_beam}, we instead perform post-expansion seeding \cite{Heinzen_2023_Cold_Atom_Source}. This enables us to take full advantage of the high beam brightness of pure helium expansions while limiting heating on the helium jet by the seeded species as well as condensation on the nozzle. 

\begin{figure}[t]
\includegraphics[width= 0.95\columnwidth]{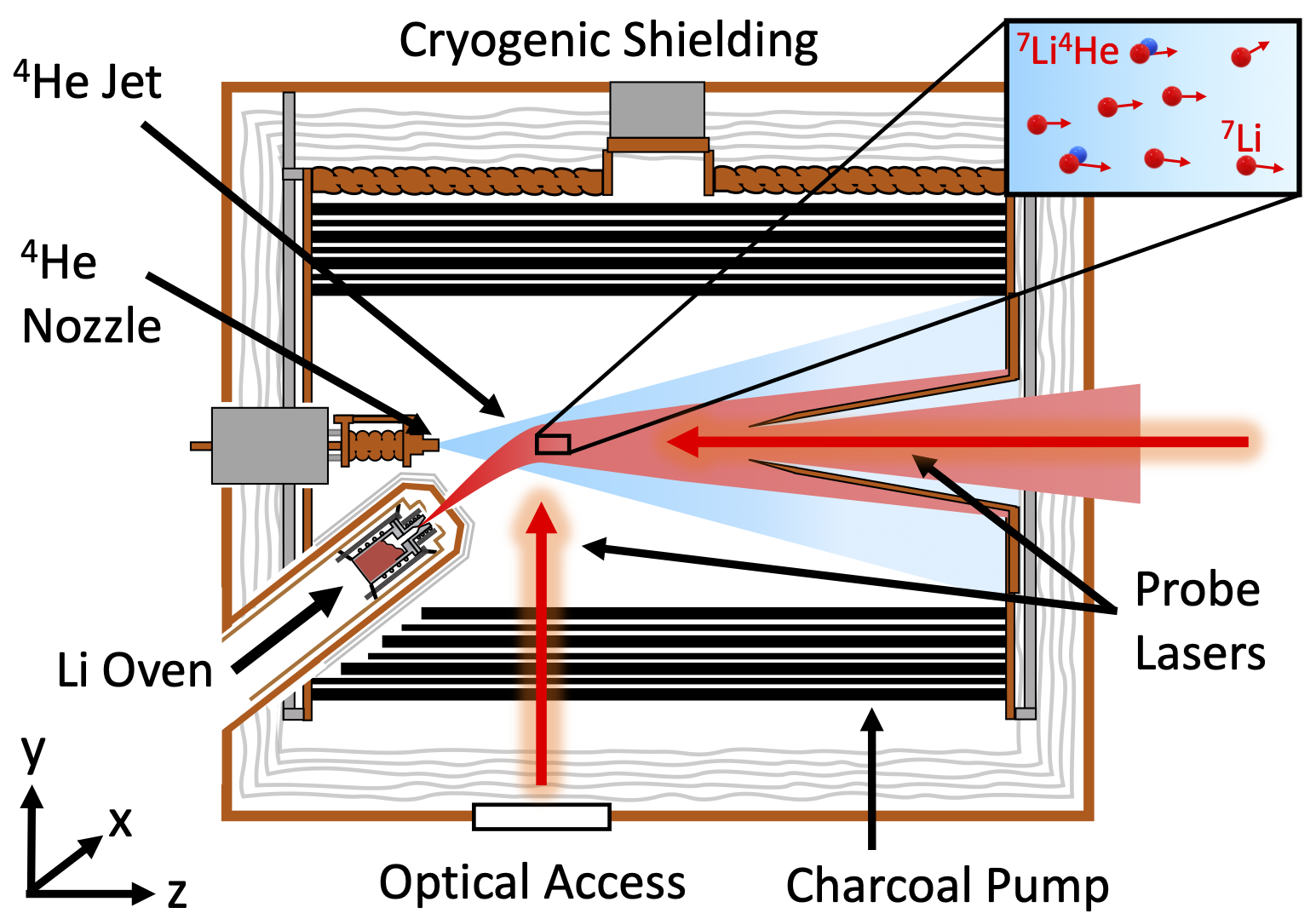}
   \caption{Experimental apparatus. An effusive beam of lithium is injected post-nozzle into a supersonic helium expansion. $^7$Li$^4$He molecules form as the lithium cools and becomes entrained in the jet. Inside the cryogenic housing a charcoal adsorption pump surrounds the helium jet. A camera collects fluorescence in the x-direction that is induced by the probe lasers.}
    \label{appratus}   
\end{figure}

\begin{figure}[t]
\includegraphics[width= 0.9\columnwidth]{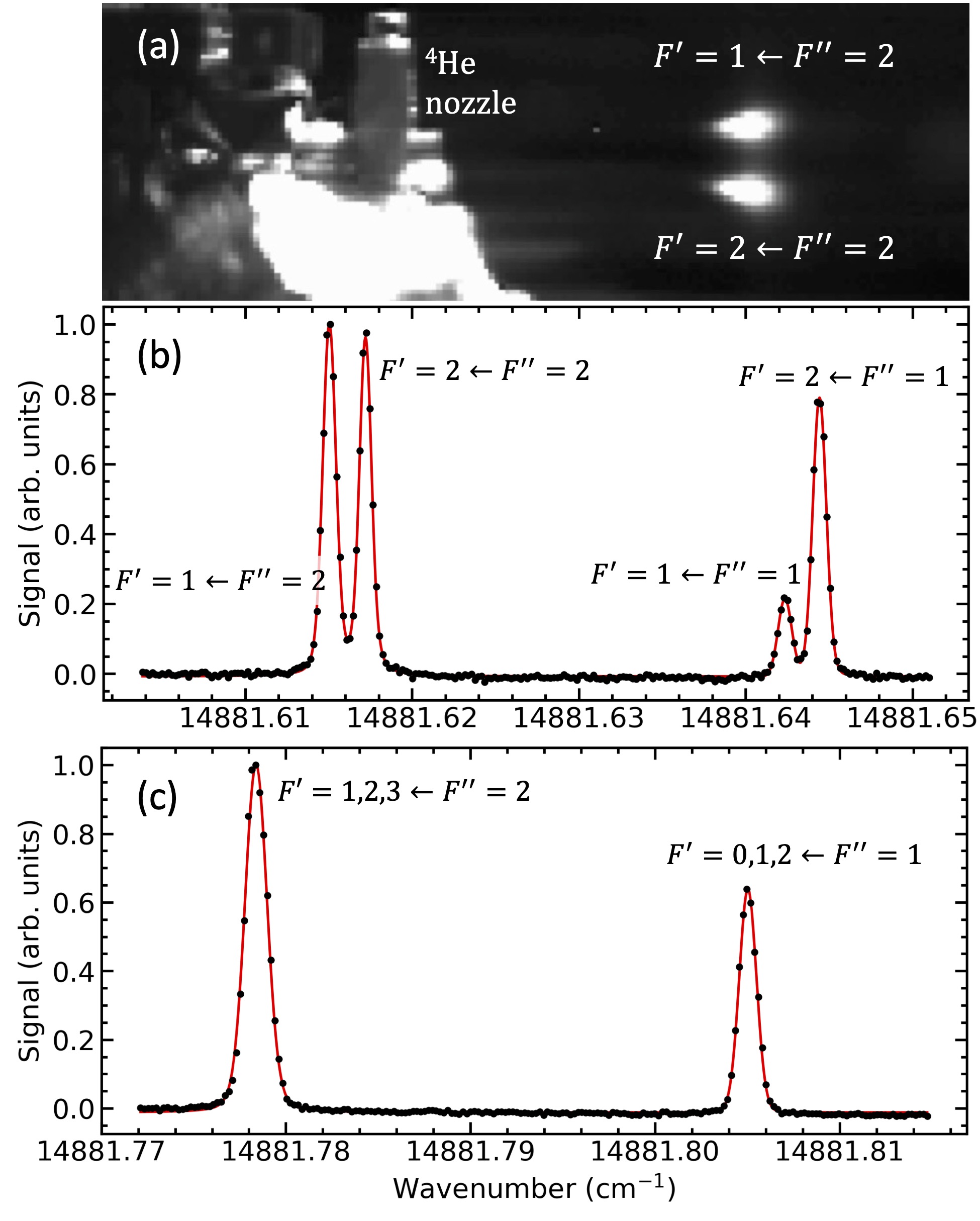}
   \caption{(a). Fluorescence image of $^7$Li$^4$He dimers at a laser frequency halfway between the $\rm{F}'=1\leftarrow \rm{F}''=2$ and $\rm{F}'=2\leftarrow \rm{F}''=2$ hyperfine transition frequencies for the  $A^2\Pi_{1/2}(\rm{v}'=5) \leftarrow X^2\Sigma^+(\rm{v}''=0)$ transition. The laser is set perpendicular to the center-line of the spherically expanding beam with two dots shown in the image. These correspond to the two hyperfine transitions separated in space as a result of Doppler shifts. (b) Spectral profile of the $A^2\Pi_{1/2}(\rm{v}'=5) \leftarrow X^2\Sigma^+(\rm{v}''=0)$ and (c) $A^2\Pi_{3/2}(\rm{v}'=5) \leftarrow X^2\Sigma^+(\rm{v}''=0)$ transitions analyzed over a region in space of height 1.35(5) mm}
    \label{spectra}   
\end{figure}

Seeding the jet with lithium, a beam of $^7$Li$^4$He is produced as the $^7$Li becomes entrained and cooled in the helium jet. The main mechanisms which form the dimers are three-body recombination $^7$Li + $^4$He + $^4$He $\rightarrow$ $^7$Li$^4$He + $^4$He and two-body exchange collisions $^7$Li + $^4$He$_2$ $\rightarrow$ $^7$Li$^4$He + $^4$He.
Laser-induced fluorescence on the single bound rovibrational ground state detects the dimers on the $A^2\Pi(\rm{v}') \leftarrow X^2\Sigma^+(\rm{v}''=0)$ transitions, with $\rm{v}' = 5 \; \rm{and} \; 6$.  
In contrast to previous work \cite{Weinstein_2013_LiHe}, the $^7$Li$^4$He dimers are 100 times colder with a longitudinal temperature of 13(6) mK. 

\begin{figure}[t]
\includegraphics[width= 0.9\columnwidth]{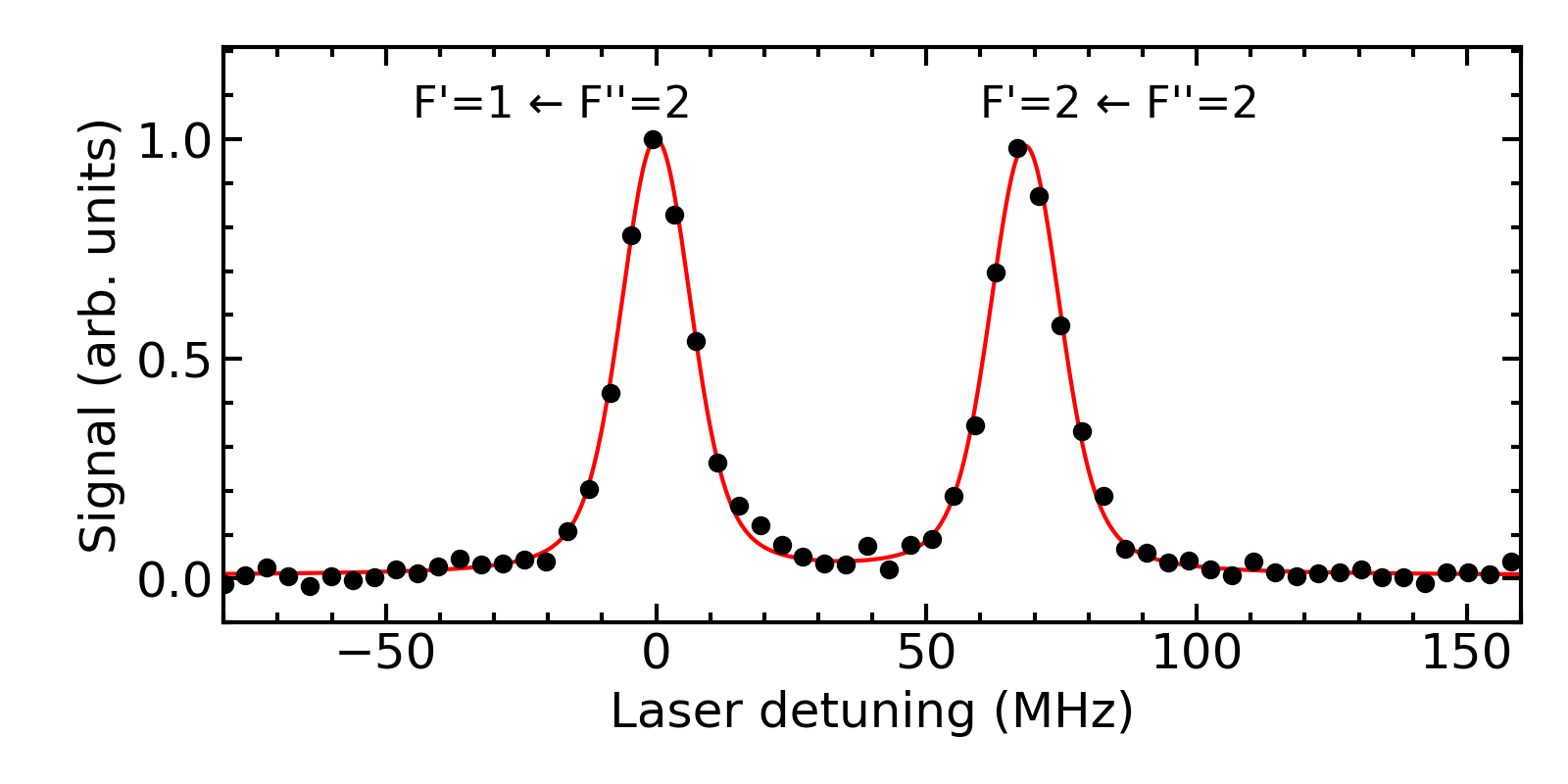}
   \caption{Fluorescence spectra of two of the hyperfine transitions for the  $A^2\Pi_{1/2}(\rm{v}'=6) \leftarrow X^2\Sigma^+(\rm{v}''=0)$ transition. Results are collected with a laser perpendicular to the expanding beam and are analyzed over a region of height 0.34(2) mm at a distance of about 4 cm from the nozzle.}
    \label{transverse_temp}   
\end{figure}

A diagram of the experimental apparatus is shown in Fig.~\ref{appratus} with further details described in Ref~\cite{Heinzen_2023_Cold_Atom_Source}. To summarize, a supersonic helium expansion is produced by flowing helium gas through a nozzle cooled to about 4 K. The cryogenic temperature of the nozzle results in a modest forward velocity of the helium beam of about $210-220$ m/s. Beneath the helium nozzle, an oven produces an effusive beam of lithium which is injected into the expanding jet.  The lithium cools via collisions with the helium. A charcoal adsorption pump cooled to around 4 K surrounds the injection region and is used to remove background helium atoms. This allows for operation with continuous helium flow rates in excess of $10^{20}$ atoms/s along with continuous lithium flow rates into the jet of over $10^{15}$ atoms/s.

Fluorescence spectroscopy is used to characterize the cooling process of both the lithium and the dimers. This is performed with an external cavity diode laser which is locked to a low finesse etalon and has a mode-hop-free tuning range of about 6 GHz. An 871A Bristol wave-meter is used for frequency calibration of the molecular transitions. The wave-meter itself is calibrated against a reference fluorescence signal of the D1 and D2 lines of lithium produced in a separate beam source. The calibration and stability of the wave-meter currently limit the accuracy with which we can determine the absolute molecular transition frequencies. 

\begin{table}[b]
\centering
\begin{tabular}{lc|c} 
\hline \hline
 &$\text{v}'=6 \leftarrow \text{v}''=0$  & $\text{v}'=5 \leftarrow \text{v}''=0$ \\ \hline
$J'=1/2 \leftarrow J''=1/2$  & & \\
\hspace{5mm}$F'=1 \leftarrow F''=2$ & 14902.561 & 14881.615 \\
\hspace{5mm}$F'=2 \leftarrow F''=2$ & 14902.563 & 14881.617 \\
\hspace{5mm}$F'=1 \leftarrow F''=1$ & 14902.587 & 14881.642 \\
\hspace{5mm}$F'=2 \leftarrow F''=1$ & 14902.589 & 14881.644 \\ \hline 
$J'=3/2 \leftarrow J''=1/2$  & & \\
\hspace{5mm}$F'=1,2,3 \leftarrow F''=2$ & 14902.740 & 14881.778 \\
\hspace{5mm}$F'=0,1,2 \leftarrow F''=1$ & 14902.768 & 14881.805 \\
\hline \hline
\end{tabular}
\caption[$A^2\Pi \leftarrow X^2\Sigma^+$ transition wavenumbers.]{$A^2\Pi \leftarrow X^2\Sigma^+$ transition wavenumbers. All values are in units of cm$^{-1}$ for vacuum with uncertainties of $\pm 0.003$ cm$^{-1}$.}
\label{table:wavenumbers}
\end{table}

\begin{figure}[t]
\includegraphics[width= 0.9\columnwidth]{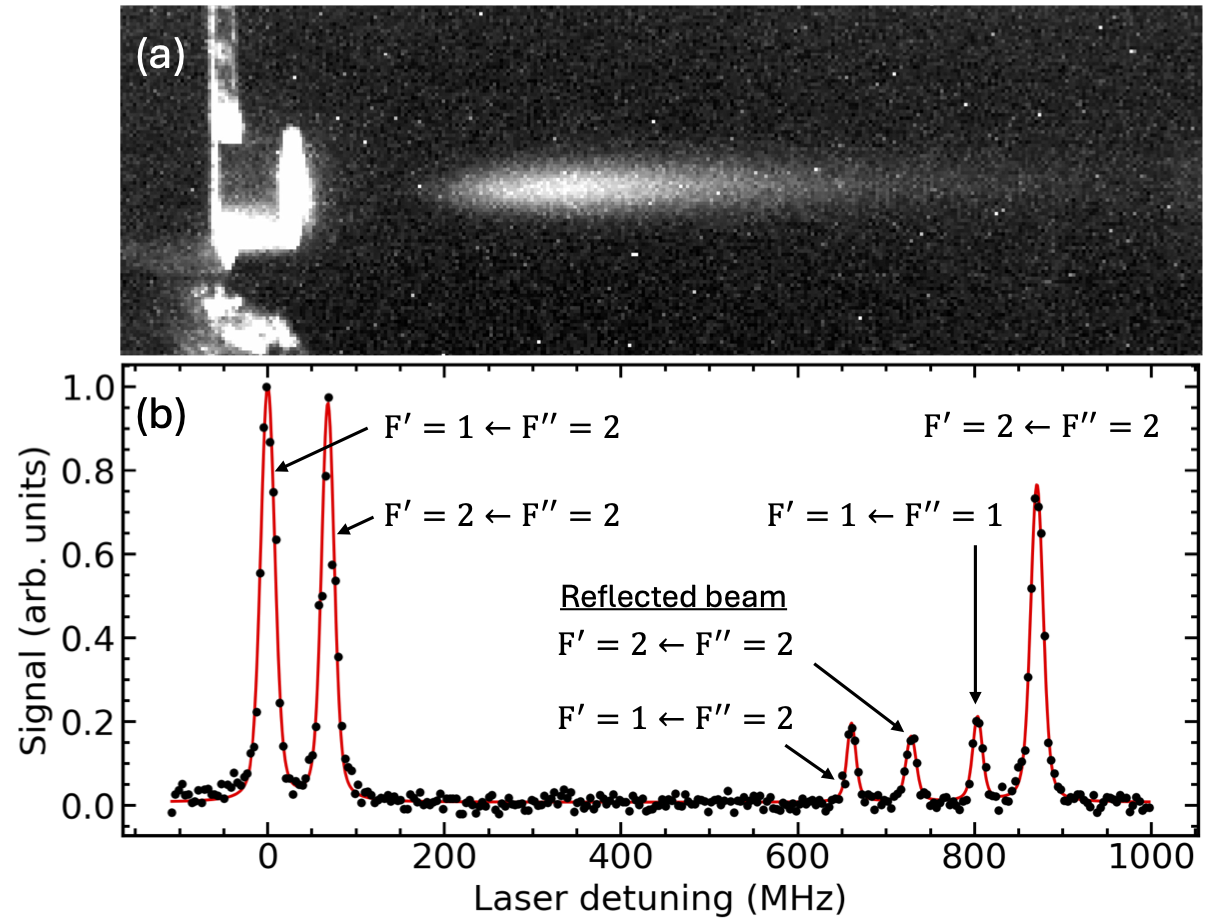}
   \caption{(a) Longitudinal fluorescence image of the dimers on the $\rm{F}'=2\leftarrow \rm{F}''=2$ hyperfine transition of the $A^2\Pi_{1/2}(\rm{v}'=6) \leftarrow X^2\Sigma^+(\rm{v}''=6)$ transition. The laser is set parallel to the center-line of the expanding beam. (b) Longitudinal spectral profile of the $A^2\Pi_{1/2}(\rm{v}'=6) \leftarrow X^2\Sigma^+(\rm{v}''=0)$  transitions. Laser detuning is from the $F'=1\leftarrow F''=2$ transition. Additional peaks are present due to reflections of the laser light off of the helium nozzle.}
    \label{longitudinal}   
\end{figure}

An example fluorescence image along with spectral profiles of the $A^2\Pi(\rm{v}'=5) \leftarrow X^2\Sigma^+(\rm{v}''=0)$ transitions are shown in Fig.~\ref{spectra} for a distance of about 4 cm from the $^4$He nozzle with a probe laser perpendicular to the expanding beam.  Absolute transition wavenumbers for both excited state transitions are given in Table~\ref{table:wavenumbers}. Due to the narrow velocity distribution in the moving frame of the jet, the $J'=1/2 \leftarrow J''=1/2$ hyperfine transitions are resolvable. As the molecular beam expands spherically and as the probe laser is perpendicular to the center-line of the expansion, there are spatially dependent Doppler shifts. The hyperfine transitions can thus be resolved as a function of frequency within a small region in space or at a single laser frequency as shown in Fig.~\ref{spectra} (a). In Fig.~\ref{transverse_temp} we show spectra over a smaller region in space of height 0.34 mm where geometric broadening is heavily reduced. Fitting the spectral profiles (black dots) in Fig.~\ref{transverse_temp} to Voigt profiles (red curves) allows us to determine the local temperature of the dimers in the moving frame of the helium jet. At a distance of 4 cm from the nozzle, this temperature is 11(4) mK. In this analysis, it is assumed that the $^7$Li$^4$He transitions have the same natural line-width of 5.87 MHz as the D1 line of $^7$Li \cite{NIST_2023_LithiumLines}. 
We find that the separation of the excited state $A^2\Pi(\rm{v}')$ with $\rm{v}' = 5 \; \rm{and} \; 6$ hyperfine transtions between $\rm{F}'=2$ and $\rm{F}'=1$ is 70(5) MHz in contrast with the $^7$Li $2^2$P$_{1/2}$ hyperfine splitting of $\sim 92$ MHz \cite{Porto_2011_Li_D1_splitting}. The ground state $X^2\Sigma^+$ hyperfine splitting between $F''=2$ and $F''=1$ is found to be 810(13) MHz, consistent with the $^7$Li ground state hyperfine splitting of $\sim 803$ MHz \cite{Porto_2011_Li_D1_splitting}. 

Longitudinal spectral profiles at a distance of about 3 cm from the nozzle are shown in Fig.~\ref{longitudinal}. 
For longitudinal results, photo-dissociation effects as well as additional noise from scattered laser light reduces our ability to resolve signal at further distances. In addition it reduces the signal to noise compared to transverse spectra. However, at this distance, the longitudinal temperature is found to be 13(6) mK. Two additional peaks are shown in Fig.~\ref{longitudinal} compared to Fig.~\ref{spectra}. This is due to reflections of the laser light off of the helium nozzle. The longitudinal probe beam produces spectral features which are blue detuned from resonance as the atoms are moving opposite to the laser propagation direction while the reflected probe beam produces spectral features which are red detuned. The additional peaks are the red detuned $F'=1 \leftarrow F''=2$ and $F'=2 \leftarrow F''=2$ transitions. Using the halfway point between the blue and red shifted Doppler features we estimate that the mean forward velocity of the dimer beam is 220(8) m/s.

From the transverse fluorescence measurements, the peak density $n_0$ of the molecular beam at a given distance from the nozzle is calculated as,
\begin{equation}
    \label{peakDensity}
    n_0 = \frac{\hbar \omega_0}{\sigma_{\rm v'v''}(\omega_0)} \frac{\Phi_{\rm p}}{ \int \La(\vec{r}) I(\vec{r}) dV} \,,
\end{equation}
where $\La(\vec{r})$ is the spatial profile of the molecular beam normalized to a height of one, $I(\vec{r})$ is the laser intensity, $\sigma_{\rm v'v''}(\omega_0)$ is the resonant Doppler broadened cross-section for transitions between vibrational levels $\rm{v'}$ and $\rm{v''}$, and $\Phi_{\rm p}$ is the number of scattered photons per second at resonance. The spatial profile is determined by fitting the fluorescence signal assuming cylindrical symmetry. The highest peak density occurs at the closest distance we can measure to the nozzle with a value of $\SI{1.8(3)e8}{}$ cm$^{-3}$ at a distance of $3.15(14)$ cm.

Evaluation of $\sigma_{\rm v'v''}(\omega_0)$ in Eq.~\ref{peakDensity} requires that we compute the Franck-Condon factors (FCF). We do so by numerically solving the radial Schr{\"o}dinger equation for the ground and excited state potentials using the discrete variable representation (DVR) approach \cite{Light_1982_DVR}. For the ground state potential, we use the $V^{\infty Z}$ potential from Ref.~\cite{Tiesinga_2020_HeLi_theory}. The excited state potential is treated as a Buckingham potential 
\begin{equation}
    V(r)=C_1e^{-C_2 r}-\frac{C_6}{r^6}\,,
    \label{potential}
\end{equation}
where $C_6 = 2.706\times10^{-78}$ J m$^{6}$ is taken from Ref.~\cite{Zhu_2001_C6}.
The coefficients $C_{1}$ and $C_{2}$ are determined by finding the values that reproduce the experimentally measured vibrational binding energies. In our fitting procedure, we neglect the fine, hyperfine, and rotational structures as these splittings are small compared to the vibrational binding energies. The optimal values of the coefficients are found to be $C_{\rm1} = 2.269\times10^{-16}$ J and $C_{\rm 2} = 2.134\times10^{-11}$ m.

All density measurements are acquired on the $A^2\Pi(\rm{v}'=6)\leftarrow X^2\Sigma(\rm{v}''=0)$ transition as it has the largest calculated FCF of $\approx0.17$. However, with this small of a FCF, most molecules will photo-dissociate upon absorption of a single photon. As a consequence, the laser intensity is kept sufficiently weak such that the probability of absorption of a photon as the molecules travel through the laser is $\ll 1$. 
\begin{figure}[t]
\includegraphics[width= 0.95\columnwidth]{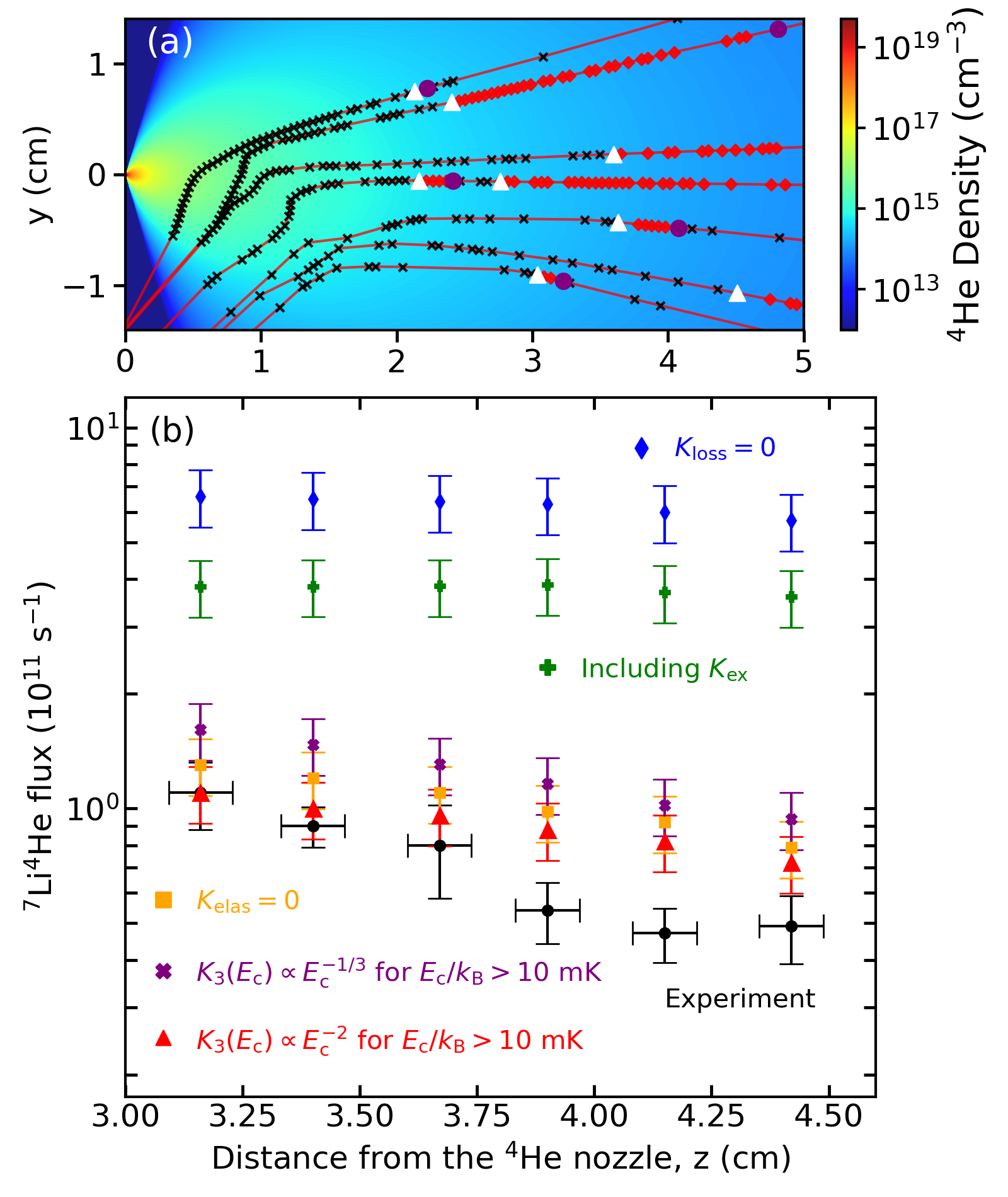}
   \caption{(a). Example particle trajectories (red curves) with the locations of $^7$Li-$^4$He collisions denoted with a black ``x", three-body formation events denoted with a white triangle, two-body loss events  denoted with a purple circle, and $^7$Li$^4$He-$^4$He elastic collisions denoted with a red diamond. Trajectories are shown on-top of the density profile of the helium jet at $x=0$. Only trajectories that lie within 0.5 mm of the $x=0$ plane are shown. Additionally, the displayed particles are selected to have minimal overlap in their trajectories. (b). Simulated and experimental $^7$Li$^4$He flux within a 2.54 cm diameter circle centered at the nozzle opening $(y,x)=(0,0)$. Results for five variations of the simulation are shown.}
    \label{simFlux}   
\end{figure}
The flux of dimers through a given area $A$ is computed as
\begin{equation}
    \Phi_{\rm dimers} = v_0 n_0 \int \La(\vec{r}) dA \,,
    \label{flux}
\end{equation}
where $v_0$ is the mean forward velocity of the molecular beam. The flux of dimers within a 2.54 cm diameter circle versus distance from the nozzle is shown in Fig.~\ref{simFlux}. Integrating over the entire spatial distribution of the beam yields a total flux of $\approx 2\times 10^{11}$ s$^{-1}$. Based on measurements of the lithium flux entering the helium jet, about $0.02\%$ of the injected lithium atoms form into dimers. The 
molecules passing through a 2.54 cm diameter circle are confined to a solid angle 
\begin{equation}
\Delta \Omega \approx \frac{\pi}{4} \left(\frac{\Delta v_{\perp}}{v_{\parallel}} \right)^2 = 0.16(2)~\rm{sr}\,,
\end{equation}
at 4 cm from the nozzle. Here $\Delta v_{\perp}$ is the full width at half maximum of the transverse velocity over the entire spatial distribution of the beam and $v_{\parallel}$ is the mean forward velocity. Studies at closer or further distances than shown in Fig.~\ref{simFlux} using the transverse laser probe are currently limited by optical access in the cryogenic region of our vacuum chamber.

\begin{figure}[t]
\includegraphics[width= 0.95\columnwidth]{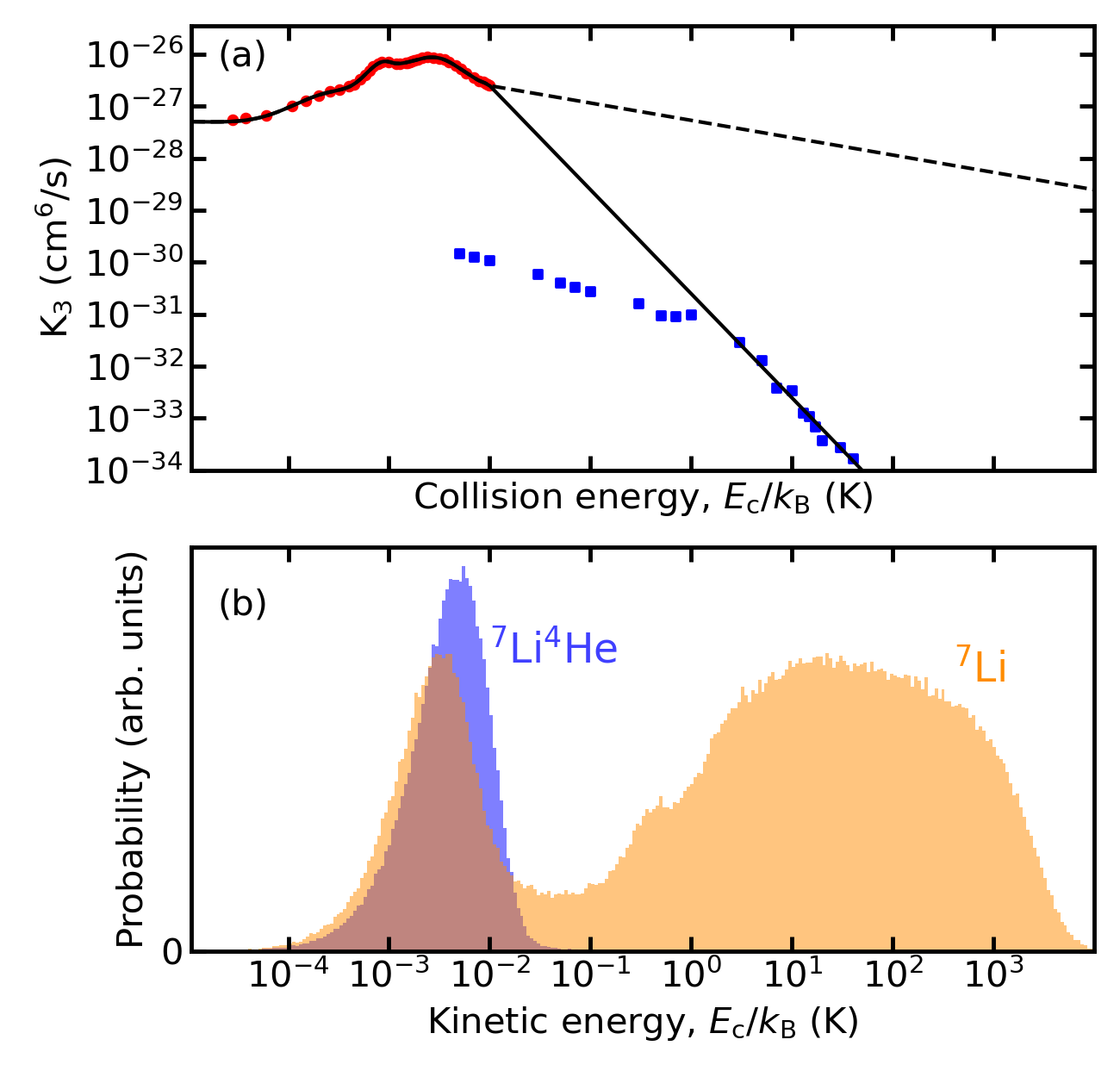}
   \caption{(a) Three-body recombination rate models (black lines) used in the Monte Carlo simulation. The solid line extending past 10 $k_{\rm B}\times$ mK corresponds to a $K_3$ value which decreases as $E^{-2}_{\rm c}$ while the dashed line decreases as $E^{-1/3}_{\rm c}$. Red circles are from Ref.~\cite{Esry_2009_recombination} with blue squares from Ref.~\cite{Mirahmadi_2021_K3_LiHe}. (b) Simulated kinetic energy distribution of the $^7$Li and $^7$Li$^4$He in the moving frame of the helium jet within a 2.54 cm diameter circle centered at $(x,y)=(0,0)$ at a distance of 4 cm from the $^4$He nozzle. The probability per fractional logarithmic energy intervals are on different scales and the heights should not be compared to each other.}
    \label{energy_dist}   
\end{figure}
A three-dimensional Monte-Carlo simulation is performed to compute the expected dimer flux. The simulation expands upon the method described in Ref.~\cite{Glick_2024_seedingSimulation}. In the simulation, the velocity phase-space probability distribution of the helium jet is modeled from theory \cite{miller_1988_free_jet, Pauly_2000_beams_book}, while the trajectories of the $^7$Li, $^7$Li$^4$He, and formation and dissociation of the $^7$Li$^4$He are treated with a Monte Carlo approach. Particles move in straight lines with time steps that are a small fraction of the local elastic collision rate, three-body recombination rate, two-body exchange rate, or two-body loss rate. At each time step $^7$Li atoms and $^7$Li$^4$He dimers may undergo two-body elastic collisions with $^4$He, $^7$Li atoms may undergo three-body inelastic collisions or two-body exchange collisions resulting in the formation of $^7$Li$^4$He dimers, and $^7$Li$^4$He dimers may undergo two-body inelastic collisions resulting in an unbound $^7$Li atom.

The rate equation governing the formation of $^7$Li$^4$He dimers is
\begin{eqnarray}
   \frac{{\rm d}n_{\rm LiHe}}{{\rm d}t} &=& \frac{1}{2!}K_{3}(E_{\rm c})n^2_{\rm He}n_{\rm Li} + K_{\rm ex}(E_{\rm c})n_{\rm He_2}n_{\rm Li}  
   \label{eq:lineshape}\\
    && - K_{\rm loss}(E_{\rm c})n_{\rm He}n_{\rm LiHe}  \,,
        \nonumber
\end{eqnarray}
where $K_3(E_{\rm c})$, $K_{\rm ex}$ and $K_{\rm loss}(E_{\rm c})$ are the three-body formation rate, two-body exchange rate and two-body loss rate respectively at a relative collision energy $E_{\rm c}$. Theoretical quantum calculations for $K_3$ for collision energies up to 10 $k_{\rm B} \times$ mK have been performed by the authors of Ref.~\cite{Esry_2009_recombination}. For large collision energies, we have examined two approaches. The first takes classical calculation from Ref.~\cite{Mirahmadi_2021_K3_LiHe} which should be reliable above $\approx 1$ $k_{\rm B} \times$ K. Modelling $K_3$ to decrease at a rate of $E_{\rm c}^{-2}$ between 10 $k_{\rm B} \times$ mK and 1 $k_{\rm B} \times$ K, smoothly connects these two regimes as shown in Fig.~\ref{energy_dist} (a). This approach does, however, likely underestimate $K_3$ at lower collision energies. In the second approach, we model $K_3$ to decrease at a rate of $E_{\rm c}^{-1/3}$ past 10 $k_{\rm B} \times$ mK in accordance with the classical threshold law found in Ref.~\cite{Greene_2014_k3_semiClassical, Mirahmadi_2021_K3_LiHe}. For $K_{\rm loss}$, we make use of theoretical calculations from Ref.~\cite{Shalchi_2020_K2} which include the processes $^7$Li$^4$He + $^4$He $\rightarrow$ $^7$Li + $^4$He + $^4$He and $^7$Li$^4$He + $^4$He $\rightarrow$ $^7$Li + $^4$He$_{2}$. The exchange rate $K_{\rm ex}$ for $^7$Li + $^4$He$_{2}$ $\rightarrow$ $^7$Li$^4$He + $^4$He is also taken from Ref.~\cite{Shalchi_2020_K2}.

The two-body $^7$Li-$^4$He elastic collision rate is determined using calculated fully quantum differential cross-sections \cite{Tiesinga_2020_HeLi_theory, Tiesinga_2021_CrossSection}.  For $^7$Li$^4$He-$^4$He collisions, we make use of the elastic rate constant $K_{\rm elas}$ from Ref.~\cite{Shalchi_2020_K2} and assume isotropic scattering. We note that the theoretically calculated rate constants from Ref.~\cite{Tiesinga_2021_CrossSection, Shalchi_2020_K2, Esry_2009_recombination, Mirahmadi_2021_K3_LiHe} use slightly different models for the ground state potential. At the moment, we are unable to quantify how these variations in the ground state potential affect the calculated rate constants and thus results from the simulations. 

\begin{figure}[t]
\includegraphics[width= 0.9\columnwidth]{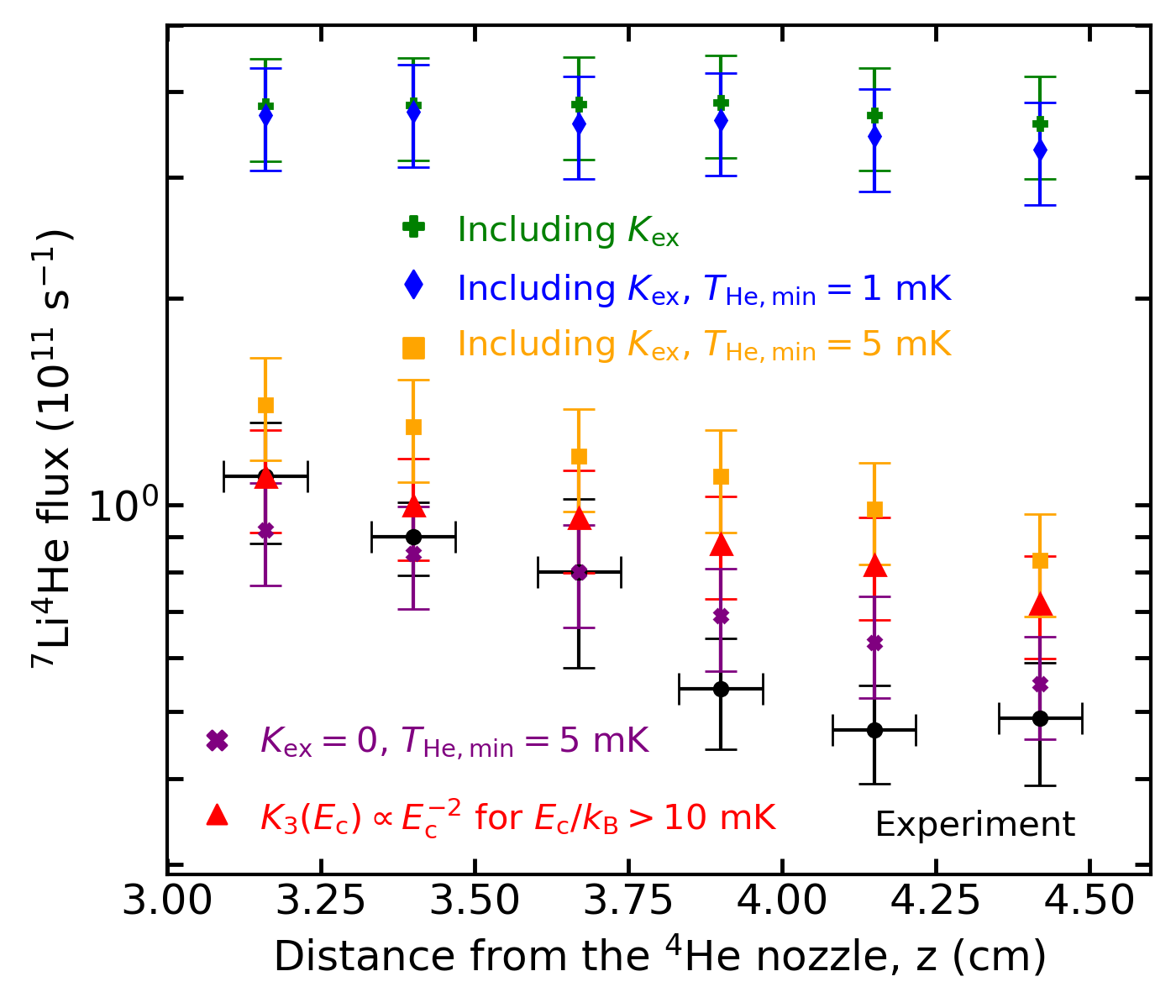}
   \caption{Simulated and experimental $^7$Li$^4$He flux within a 2.54 cm diameter circle centered at the nozzle opening for models where an artificial temperature floor is set for the helium expansion.}
    \label{simFluxTempFloor}   
\end{figure}

We treat scattering for three-body inelastic collisions and two-body collisions which remove $^7$Li$^4$He as isotropic with momentum and energy being conserved. We assume the binding energy is $34$ $k_{\rm B}\times$ mK based on the measured binding energy of $34(36)$ $k_{\rm B}\times$ mK from Ref.~\cite{Weinstein_2013_LiHe}. This sets the amount of kinetic energy added or removed during the collisions. Shown in Fig.~\ref{simFlux}(a), this small change in kinetic energy amounts to a negligible change in particle's trajectories post-collision. In fact, we find that varying the kinetic energy added by a factor of two results in just a few percent change in the simulation flux. As such, for computational simplicity, all loss collisions are treated as $^7$Li$^4$He + $^4$He $\rightarrow$ $^7$Li + $^4$He + $^4$He to determine the post-collision velocity of the $^7$Li atoms. 

Simulated kinetic energy distributions of the $^7$Li and $^7$Li$^4$He in the moving frame of the jet are presented in Fig.~\ref{energy_dist} (b). In contrast with the energy distribution of the $^7$Li$^4$He, the $^7$Li distribution indicates a significant portion of $^7$Li atoms that have not fully thermalized with the helium. This behavior is observed experimentally in Ref.~\cite{Heinzen_2023_Cold_Atom_Source} with fluorescence measurements showing an asymmetric Doppler profile indicating a sizable fraction of lithium atoms that have not fully thermalized. Notably, the dimer fluorescence, as shown in Fig.~\ref{spectra} has no observable asymmetry. These results can be understood by examining the energy-dependent three-body recombination rate shown in Fig.~\ref{energy_dist} (a) alongside the energy distribution of the $^7$Li atoms. The portion of $^7$Li atoms with high kinetic energies simply have a drastically reduced probability of undergoing three-body collisions. 

As discussed below, we are currently unable to accurately determine the density of helium dimers in our system and their contribution to the formation of $^7$Li$^4$He dimers. We have designed our source, as discussed in Ref.~\cite{Heinzen_2023_Cold_Atom_Source}, to operate in a regime where helium cluster formation should be minimal. However, as exchange collisions are two-body process, this can be an important mechanism even with a $^4$He$_2$ density that is a small fraction of the $^4$He density. 
We therefore explore a number of models to examine how variations in different rate constants affect the simulated dimer flux. Results from five versions of our simulations are shown Fig.~\ref{simFlux}. The ``baseline" version of the simulation is the red triangles in Fig.~\ref{simFlux} (b) in which $K_{\rm ex} = 0$. Other markers corresponding to changes in a single rate constant of the baseline version. For example, setting the loss rate to zero results in $\sim$ 7 times higher flux while removing $^7$Li$^4$He-$^4$He collisions results in $10-20\%$ higher flux. The two approaches for modeling $K_3$ for $E_{\rm c} > 10$ mK, vary by $20-50\%$. Agreement with experiment is highest when including $^7$Li$^4$He-$^4$He collisions, dissociation collisions, and modeling $K_3$ to decrease at a rate of $E_{\rm c}^{-2}$ past 10 $k_{\rm B} \times$ mK. For this  model, the average fractional difference in flux between experiment and simulation is $0.36(30)$. When modeling $K_3$ to decrease as $E_{\rm c}^{-1/3}$ past 10 $k_{\rm B} \times $ mK, the average fractional difference is 0.83(37). 

Using a rate equation model discussed in the Appendix, we have estimated that the helium dimer density could be as high as $\sim0.6\%$ of the helium density. This assumes a pure adiabatic expansion where heat of condensation contributes to negligible heating of the helium. Modelling the helium dimers as being in thermal equilibrium with the helium, we find that the inclusion of exchange collisions results in a 4-6 times increase in the $^7$Li$^4$He flux over the range of distances that we measure. This would indicate that this is the dominant process over three-body formation. However, unlike the capture efficiency of $^7$Li and three-body formation of $^7$Li$^4$He, the density of $^4$He${_2}$ dimers depends strongly on minor increases in the helium temperature on the order of a few milliKelvin. This is due to the helium binding energy being on the order of 1 $k_{\rm B} \times$ mK, while the $^7$Li$^4$He binding energy is on the order of 30 $k_{\rm B} \times$ mK. Some results discussed in Ref.~\cite{Heinzen_2023_Cold_Atom_Source}, show that minor temperature increased in the model of the helium temperature yield better agreement with experimental results of the seeded $^7$Li. Using a simple model in which a temperature floor is set in the helium expansion we repeat our simulations with a temperature floor of 1 mK and 5 mK. Results are shown in Fig.~\ref{simFluxTempFloor}. A temperature floor of 1 mK reduces the $^7$Li$^4$He flux by about 10$\%$ when including exchange collisions and only by a few percent (not shown) when $K_{\rm ex} = 0$. For a temperature floor of 5 mK, the flux including exchange collisions decreases by a factor of 3 to 4 while the flux with $K_{\rm ex} = 0$ decreases by at most 20 $\%$. While this model does not capture the full dynamics of heating of the helium jet, it indicates that the contributions from exchange collisions have a strong temperature dependence. 

Without a more physical model of our system that includes helium dimer formation, future studies will be necessary to determine quantitative rate constants. Monte-Carlo simulations of the entire helium expansion, helium dimer formation, and seeding of the lithium and formation of lithium-helium is one path forward but is prohibitive for our current computational resources. Experimentally, the helium dimer density could be determined with a transmission grating \cite{Toennies_2002_HeDimers} while the helium temperature could be accessed using a pulsed electron beam to make meta-stable helium and performing time of flight measurements. We note that our apparatus was not specifically designed to produce a high flux of $^7$Li$^4$He dimers or $^4$He$_{2}$ dimers or to allow for measurements over a very wide range of temperatures and densities. A modified apparatus could produce a higher flux of dimers and trimers over a wider range of densities and temperatures and enable studies of three-body collisions as well as the possiblity of exchange collisions. We hope to explore this in future work.

In summary, we have demonstrated an approach for producing a high flux beam of $^7$Li$^4$He dimers with a temperature on the order of 10 mK. This method should be well suited for producing a number of ground state alkali-helium pairs given that theory predicts comparable recombination rates as $^7$Li + $^4$He + $^4$He $\rightarrow$ $^7$Li$^4$He + $^4$He \cite{Esry_2009_recombination}.  Furthermore, all alkali-helium pairs have a large collisional cross-section with helium at low collision energies which is necessary for cooling and entrainment in the jet \cite{Kleinekath_1999_singleBoundState}. For example, in Ref.~\cite{Glick_2024_seedingSimulation}, we simulate the seeding process for $^{87}$Rb and predict that $\sim0.1\%$ of injected rubidium atoms could be entrained within a 0.02 sr solid angle with an energy of less than $10$ mK in the moving frame of the helium at a distance of 4 cm. This is roughly an order of magnitude less efficient than with lithium. Even with a dimer formation rate that is an order of magnitude less than for helium-lithium, this would roughly amount to a beam flux of $10^{9}$ dimers/s. As such, our approach could facilitate quantitative studies of three-body recombination and other few-body collision processes for a variety of molecules including those yet to be observed experimentally.

We gratefully acknowledge Eite Tiesinga and Jacek K\l{}os, who shared their calculated $^7$Li-$^4$He cross sections with us. We also thank Paul Kunz for sharing lab equipment with us. We acknowledge the financial support of this work by Fondren Foundation and the Army Research Laboratory Cooperative Research and Development Agreement number 16-080-004.

\section{Appendix}
We estimate the density of helium dimers as a function of distance $r$ from the nozzle along the centerline of the expansion $(x,y) = (0,0)$ by modeling the formation process as

\begin{equation}
\begin{aligned}
        \frac{{\rm d}n_{\rm He_2}}{{\rm d}t} &= \frac{1}{3!}\langle K_{3} \rangle n^3_{\rm He} - \langle K_{\rm 2} \rangle n_{\rm He_2}n_{\rm He} - \frac{2}{t}n_{\rm He_2} \\
        \frac{{\rm d}n_{\rm He}}{{\rm d}t} &= -\frac{1}{3!}\langle K_{3} \rangle n^3_{\rm He} + \langle K_{\rm 2} \rangle n_{\rm He_2}n_{\rm He} - \frac{2}{t}n_{\rm He}.
\end{aligned} 
\label{he2_rate_eqn}
\end{equation}
Here $K_3$ and $K_2$ are the three-body recombination and two-body dissociation rates with brackets indicating a thermal average. The thermally averaged rate constants are computed using the energy-dependent recombination and dissociation rates from Ref.~\cite{Esry_2008_He2} and assuming that the helium and helium dimers are in thermal equilibrium and have a temperature which follows from Eq.~\ref{jet_temp}.

\begin{figure}[t]
\includegraphics[width= 0.95\columnwidth]{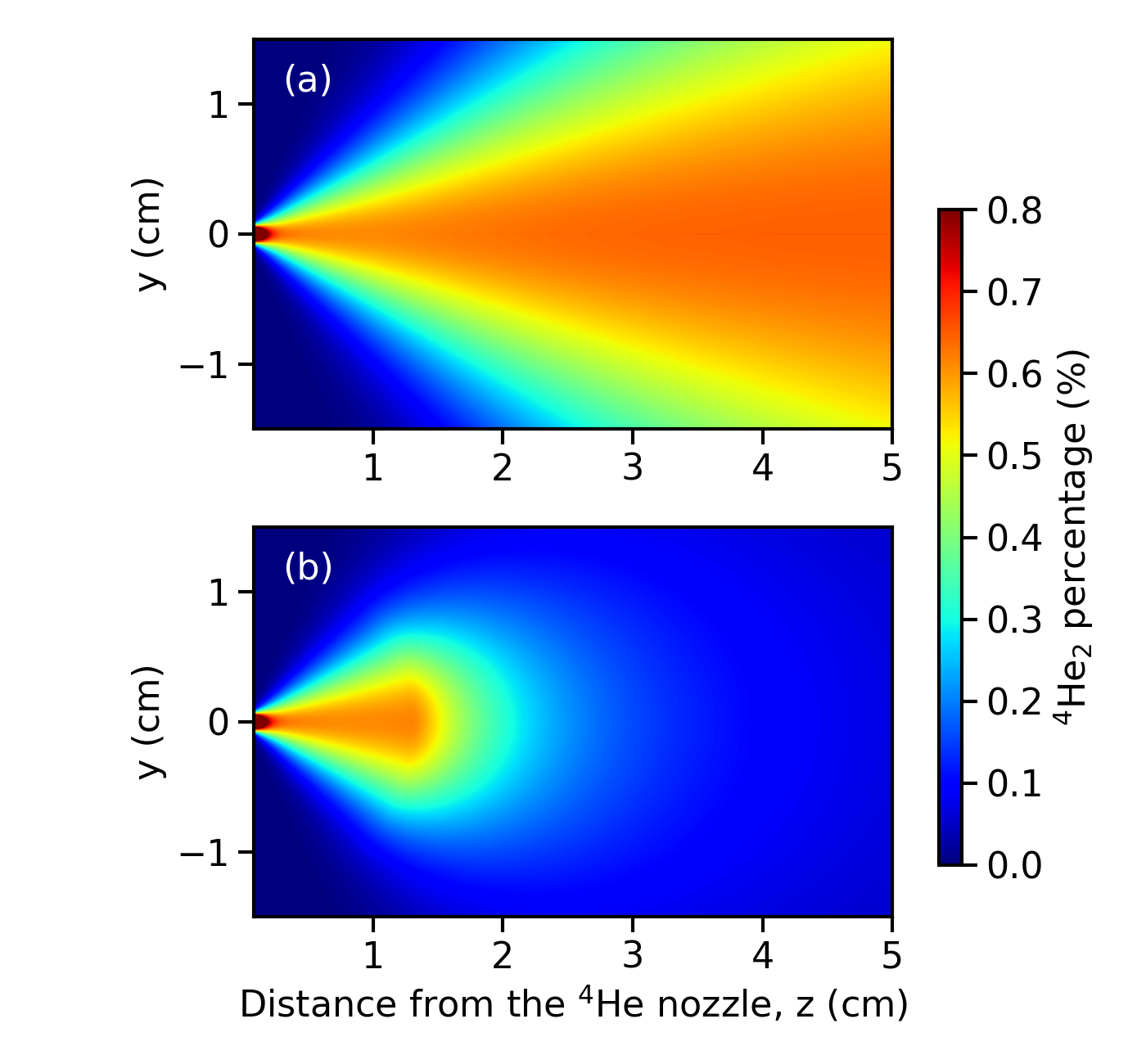}
   \caption{Simulated spatial distribution of $^4$He$_2$ as a percentage of the $^4$He density centered at $x=0$. Results for a pure adiabatic expansion are shown in (a) while (b) shows results when introducing a temperature floor of 5 mK in the expansion.}
    \label{simHe2}   
\end{figure}

In our model, we approximate the flow speed of the helium and helium dimers as being equal to the terminal velocity $v_{\rm tv}$ of the jet such that $t= r/v_{\rm tv}$. This is due to the general feature that for supersonic expansions the flow speed approaches within a few percent of the terminal velocity within about five nozzle diameters \cite{miller_1988_free_jet}. For our source, the nozzle diameter $d_0$ is 200 microns. The final terms in the rate equations account for the density decreasing as a rate of $r^{-2}$ with distance from the nozzle. For our nozzle geometry, the temperature of the helium after a distance of a few nozzle diameters is expected to decrease as
\begin{equation}
    T(r) = 0.287~T_0 \left(\frac{r}{d_0}\right)^{-4/3} \,,
    \label{jet_temp}
\end{equation}
where $T_0$ is the reservoir temperature of the helium. Eq.~\ref{jet_temp} will hold in the limit that heat of condensation from dimer formation is negligible as well as a helium-helium collision rate which is sufficient to maintain local thermal equilibrium. For our source conditions, as discussed in Ref.~\cite{Glick_2024_seedingSimulation}, we expect the helium to remain highly collisional for distances exceeding 10 cm from our nozzle. Finally, for determining the dimer density off axis, the angular dependence of the helium density for moderate angles from the center-line of the expansion is treated as 
\begin{equation}
    n_{\rm He}\left(\theta\right) = n_{\rm He} \cos^2(1.15\theta) \,.
\end{equation}
Further discussions of the properties of supersonic expansions can be found in Ref.~\cite{miller_1988_free_jet, Pauly_2000_beams_book}. 

Numerically solving Eq.~\ref{he2_rate_eqn} for our source conditions starting from four nozzle diameters yields a helium dimer density that is $\sim 0.6\%$ that of the helium past a few cm from the nozzle. The angular dependence at $x=0$ is shown in Fig.~\ref{simHe2}. Also shown in Fig.~\ref{simHe2} is the simulated helium dimer density for a model of the helium expansion where a temperature floor of 5 mK is set. Compared to a pure adiabatic expansion, this temperature floor decreased the fraction of helium dimers by about a factor of 6 at 4 cm from the nozzle. While this is not a physically accurate way to model heat of condensation or heating from the lithium, it demonstrates the strong temperature dependence on the formation of helium dimers by minor increases to the helium temperature.

\bibliography{references}

\end{document}